\documentclass[aps,prb,twocolumn,floatfix,showpacs,citeautoscript,longbibliography,hyperlinks]{revtex4-2}

\usepackage{graphicx}
\usepackage{bm}
\usepackage{hyperref}

\begin{document}

\title{Photonic heat transport from weak to strong coupling}
\author{Minh Tam}
\author{George Thomas}
\author{Dmitry S. Golubev}
\affiliation{Pico group, QTF Centre of Excellence, Department of Applied Physics, Aalto University, 00076 Aalto, Finland}

\begin{abstract}
Superconducting circuits provide a favorable platform for quantum thermodynamic experiments. 
An important component for such experiments is a heat valve, i.e. a device which allows one to
control the heat power flowing through the system.    
Here we theoretically study the heat valve based on a superconducting quantum interference device (SQUID) coupled 
to two heat baths via two resonators. The heat current in such system
can be tuned by magnetic flux. We investigate how does the heat current modulation depend on 
the coupling strength $g$ between the SQUID and the resonators. 
In the weak coupling regime the heat current modulation grows as $g^2$, but,
surprisingly, at the intermediate coupling it can be strongly suppressed.  
This effect is linked to the resonant nature of the heat transport at weak coupling, where
the heat current dependence on the magnetic flux is a periodic set of narrow peaks. At the intermediate
coupling the peaks become broader and overlap, thus reducing the heat modulation. 
At very strong coupling the heat modulation grows again and finally saturates at a constant value.
\end{abstract}

\maketitle

\section{Introduction}
  
Quantum thermodynamics attracts a lot of attention 
both from the fundamental physics viewpoint and due to potential applications in nanoscale devices \cite{Deffner2019,Strasberg2021,Myers2022}. 
In this context, understanding of the heat transport in nanoscale systems  is very important
\cite{Dhar,Dhar2006,Dubi2011,Li2012,Mosso2017, Photonic2019, pekolareview2021}.
Precise control and tuning of the heat power  is essential for the design of quantum heat engines \cite{Scovil,Abah2012,Uzdin2015,Benenti2017,Brandner2017,Campisi2016,Diaz2021}, thermal rectifiers\cite{Segal2005,Ruokola2009,Bhandari2021,Goury2019}, transistors \cite{Julian2016,Majland2020}, masers \cite{Thomas2020} and circulators \cite{Diaz2021}. 
Such thermal devices can also be used for heat management in quantum circuits \cite{pekolareview2021,Pekola2015}. 
In superconducting circuits one can control the heat current by tuning the transition frequencies of a qubit 
\cite{Ronzani2018,Photonic2019,Xu2021} and 
very accurately measure it employing, for example, normal metal - insulator - superconductor junctions as thermometers \cite{Giazotto2006}. 
The heat transport experiments in superconducting circuits can be performed at very low temperatures, 
where the photonic heat flux dominates over phononic and electronic contributions \cite{Schmidt2004}. 
In such systems, the heat can be transmitted over macroscopic distances \cite{Partanen2016},
which permits remote  management of the heat.

Thermodynamics of the  systems weakly coupled to the environment has been studied extensively, 
and there also have been many extensions of the theory to the strong coupling regime \cite{Segal2006,Nicolin2011,Wang2015,He2018,Xu2016}. 
One of the difficulties in this context is
the ambiguity in the definition of heat at strong coupling \cite{Talkner2020}. 
Here we consider a system consisting of two small normal metal islands coupled to the two coplanar waveguide resonators,
which are, in turn, coupled to the superconducting quantum interference device (SQUID) 
tunable by magnetic flux (see Fig. \ref{fig: resonator_qubit_resonator}). 
We express the heat in terms of the temperature changes of the metallic islands playing the role of thermal baths.
Thus, in our model the heat is defined via the changes of the internal energies of the baths. 
This definition is inspired by the experiments mentioned above, and
the heat defined in this way can be experimentally measured regardless of the coupling strength 
between the SQUID and the baths. 

The device depicted in Fig. \ref{fig: resonator_qubit_resonator} is supposed to operate as a heat valve, which 
allows one to tune the heat flux between the resistors by changing the critical current of the SQUID with the magnetic flux.
The performance of the valve is characterized by the heat current modulation amplitude, i.e. by the difference between the
maximum and the minimum values of the heat current. We investigate how does the heat modulation vary with various parameters and 
obtain two surprising results. First, we find that the modulation of the heat
depends on the coupling strength between the SQUID and the resonators in non-monotonous way. Indeed, 
the modulation grows with the coupling strength in the weak coupling regime, it almost vanishes at the intermediate coupling,
then it grows again and eventually saturates at very strong coupling.  
Second, the strongest heat modulation is achieved in the weak coupling regime.
In our modelling we use feasible parameters \cite{Photonic2019,Ronzani2018,Xu2021}, and we believe that our predictions can be experimentally tested.
Finally, we have derived analytical expressions for the heat flux in various limiting cases in terms of the circuit parameters. 
 
We organize the paper as follows: in Sec. II, we introduce the model and analytically analyze the weak, the intermediate and the strong coupling regimes 
and in Sec. III we summarize the results.

\section{The model}

\begin{figure}[ht!]
    \centering
    \includegraphics[width=0.7\columnwidth]{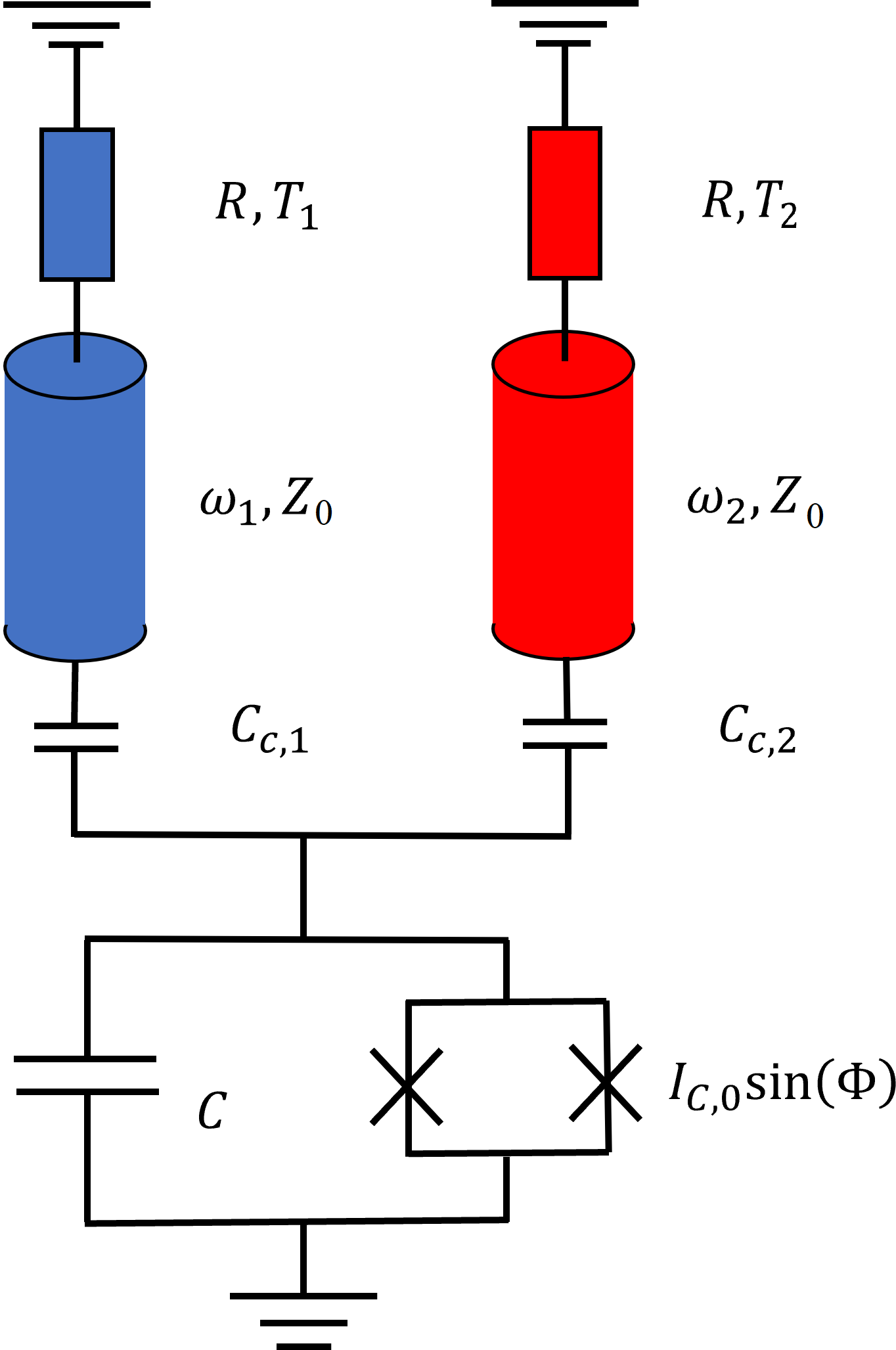}
    \caption{The schematics of the heat valve under consideration. The hot bath is indicated by red color and cold one - by blue color. }
    \label{fig: resonator_qubit_resonator}
\end{figure}

We consider an electric circuit depicted in Fig. \ref{fig: resonator_qubit_resonator}. 
In this circuit, the two normal metal islands, having the same 
resistances $R$ and kept at constant temperatures $T_{1},T_2$, act as heat baths.
The temperatures $T_{j}$ ($j=1,2$) can be experimentally monitored using biased 
normal metal - insulator - superconductor junctions \cite{Giazotto2006}. 
The two identical  superconducting
coplanar waveguide $\lambda/4$-resonators with characteristic impedance $Z_0$ serve as filters. 
The resonators are coupled to the SQUID via the capacitors $C_{c,j}$. 
The frequencies of the resonators, $\omega_1$ and $\omega_2$ may slightly differ to compensate
the difference between $C_{c1}$ and $C_{c2}$, as we explain below.

In this setup, the SQUID can act as a quantum heat valve \cite{Ronzani2018,pekolareview2021}.  
Indeed, it provides a control parameter, the external magnetic field, 
which one can use to tune the heat current through the system. 
We assume that the SQUID is symmetric and its critical current $I_C$ periodically depends on the magnetic flux $\Phi$ as 
\begin{equation}
I_{C}(\Phi)=I_{C, 0}\left|\cos \left(\pi \Phi / \Phi_{0}\right)\right|.
\end{equation}
Here $I_{C,0}$ is the critical current at zero flux and $\Phi_0$ is the magnetic flux quantum.
The SQUID is characterized by the Josephson energy $E_{J}(\Phi)=\hbar I_{C}(\Phi) / 2 e$ 
and by the charging energy $E_{C}=e^{2} / 2\left(C_{c,1}+C_{c,2}+C\right)$, where $C$ is the capacitance of the SQUID, 
see Fig. \ref{fig: resonator_qubit_resonator}. 
Here we consider the limit $E_{J}(0) \gg E_{C}$ and $E_{C}\lesssim k_BT_{j} \lesssim 2E_J(0)$. 
In this case, the two non-linear Josephson junctions of the SQUID can be approximately replaced by an inductor $L_{J}(\Phi)=\hbar/2 e I_{C}(\Phi)$, 
and the SQUID as a whole - by an $L C$-circuit with the frequency 
\begin{eqnarray}
\omega_{J}(\Phi)=\sqrt{8 E_{J}(\Phi) E_{C}} / \hbar.
\label{omega_J}
\end{eqnarray} 

To describe the transport of heat by photons between the resistors 1 and 2, we use a quasiclassical Langevin equation where 
the power spectra of Nyquist noises generated by the resistors are determined by the fluctuation-dissipation theorem \cite{Schmid1982}. 
In this way, we obtain the following expression for the heat current from the resistor 2 to the resistor 1\cite{Pascal}
\begin{equation}
    \label{eq: heat formula}
    J(\Phi)=\int_{0}^{\infty} d\omega \frac{\hbar \omega}{2 \pi} \tau(\omega,\Phi)\left[N_{2}(\omega)-N_{1}(\omega)\right].
\end{equation}
Here $N_{j}(\omega) = 1 /\left(e^{\hbar \omega / k_{B} T_{j}}-1\right)$ are the Bose functions and  $\tau(\omega,\Phi)$ is the transmission probability,
which depends on frequency and magnetic flux. 
The transmission probability $\tau(\omega,\Phi)$ equals to the square absolute value of the transmission coefficient $|S_{21}(\omega,\Phi)|^2$ 
between the two resistors. 
Eq. (\ref{eq: heat formula}) has the familiar form of the Landauer formula for the photon current \cite{Dhar}. 
For the circuit under consideration, see Fig. \ref{fig: resonator_qubit_resonator}, the transmission probability $\tau(\omega,\Phi)$ is given by \cite{Photonic2019,Pascal}
\begin{equation}
    \label{eq: tau formula}
    \tau(\omega,\Phi)=\frac{4\,{\rm Re}\left[\frac{1}{Z_{1}(\omega)}\right]\, 
{\rm Re}\left[\frac{1}{Z_{2}(\omega)}\right]}{\left|-i \omega C+\frac{1}{Z_{1}(\omega)}+\frac{1}{Z_{2}(\omega)}+\frac{1}{Z_{J}(\omega,\Phi)}\right|^{2}},
\end{equation}
where the impedances of the resonators are
\begin{eqnarray}
\label{eq: bath impedance}
Z_{j}(\omega)&=&\frac{1}{-i \omega C_{c,j}}-i Z_{0} \tan \left(\frac{\pi}{2} \frac{\omega}{\omega_j}+i \alpha\right), \\
\label{eq: alpha}
\alpha&=&\frac{1}{2} \ln \frac{Z_{0}+R}{Z_{0}-R},
\end{eqnarray}
and the impedance of the SQUID is
\begin{equation}
    \label{eq: Squid impedance}
    Z_{J}(\omega,\Phi)= -i\omega L_J(\Phi) = \frac{-i \hbar \omega}{2 e I_{C}(\Phi)}.
\end{equation}
In Eq. (\ref{eq: bath impedance}) the angular frequencies $\omega_j$ correspond to the fundamental modes of the uncoupled resonators.

\begin{figure*}[t]
    \centering
    \includegraphics[width=0.97\textwidth]{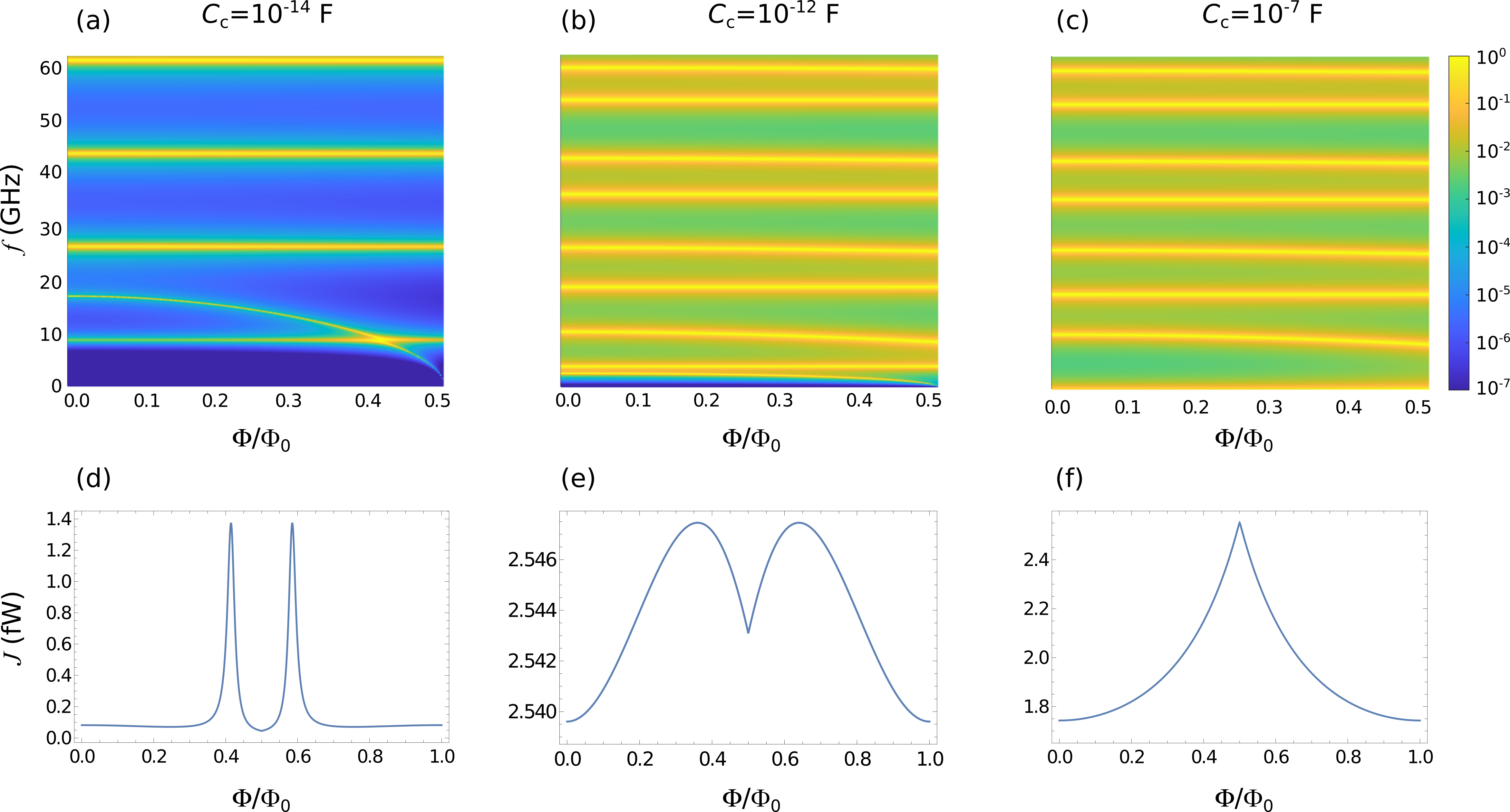}
    \caption{Panels (a),(b) and (c) --- transmission probability (\ref{eq: tau formula}) versus normalized magnetic $\Phi/\Phi_0$ and the frequency $f=\omega/2\pi$.
Panels (d),(e) and (f) --- heat current (\ref{eq: heat formula}) as a function of $\Phi/\Phi_0$.
(a,d) Weak coupling regime with $C_c=10$ fF. The corresponding value of the coupling constant (\ref{eq: g weak}) is $g/2\pi=296$ MHz, 
of the shifted resonator frequency (\ref{Omega_r}) $\Omega_r/2\pi=8.73$ GHz and of the damping rate (\ref{kappa_1}) is $\kappa/2\pi = 451$ MHz. 
(b,e) Intermediate coupling regime with $C_c=1$ pF, the coupling constant (\ref{g1}) $g/2\pi=2.592$ GHz, 
the frequency of the uncoupled mode (\ref{omega_unc}) $\omega_{\rm unc}/2\pi=4.15$ GHz, the frequency of the coupled mode (\ref{Omega_r1})
$\Omega_r/2\pi=8.425$ GHz and $\kappa/2\pi = 451$ MHz.
(c,f) Strong coupling regime with $C_c=10^{-7}$ F, $g/2\pi=2.975$ GHz, $\omega_{\rm unc}/2\pi=14.87$ MHz, $\Omega_r/2\pi=8.414$ GHz and $\kappa/2\pi = 451$ MHz.}
    \label{fig: density plot}
\end{figure*}

In Figs. \ref{fig: density plot} and \ref{fig: min_max J} we plot, respectively, the transmission probability (\ref{eq: tau formula})
and the heat current (\ref{eq: heat formula}) evaluated numerically.
For the numerical simulations we have used the parameter values typical for the experiment:
$Z_0 = 50 \Omega$, $R = 2 \Omega$, $\omega_1/2\pi=\omega_2/2\pi=8.84$ GHz, $C=58.7$ fF, $I_C = 291$ nA, $T_2 = 300$ mK, $T_1 = 150$ mK. 
In the subsequent subsections we discuss various approximate approaches, which allow us to find
analytical expressions for the heat current and to understand the underlying physics.

\subsection{Qualitative discussion}

The transmission probability (\ref{eq: tau formula}) has peaks at frequencies corresponding to the eigenmodes of 
the whole system "two resonators plus SQUID". The position, the height and the width of these peaks depend on magnetic flux.
In Fig. \ref{fig: density plot} we plot the function $\tau(\omega,\Phi)$ for three different values of the coupling
strength between the SQUID and the resonators.
In this figure and  in the rest of the paper, we assume that the maximum value of the SQUID frequency satisfies
\begin{eqnarray}
\hbar\omega_{1,2} < \omega_J(0) < 3\hbar\omega_{1,2}.
\end{eqnarray}
In this case,  the SQUID frequency (\ref{omega_J})  crosses only the lowest resonator modes.  

In the weak coupling limit (Fig. \ref{fig: density plot}a) the modes of the resonators and of the SQUID are almost independent.
They become hybridized only in the vicinity of the flux point where $\omega_J(\Phi)$ crossed the frequency of the resonators $\omega_{1,2}$.
The heat current through the system $J(\Phi)$ shows sharp peak at this point and almost vanishes away from it (Fig. \ref{fig: density plot}d). That is why 
the modulation of the heat current in the weak coupling limit approximately equals to its maximum value.

At the intermediate coupling (Fig. \ref{fig: density plot}b), the hybridization between the resonators and the SQUID becomes significant even far away from
the crossing flux point. For this reason, the heat current peaks become broad and overlap (Fig. \ref{fig: density plot}e).
Therefore, the magnitude of the heat current modulation drops. In fact, it almost vanishes, see Fig. \ref{fig: min_max J}.
Another effect, visible in Fig. \ref{fig: density plot}b, is the splitting of the resonator modes into pairs. 
In each pair, only one of the modes
is coupled to the SQUID and is sensitive to the magnetic flux. Namely, it is the mode having voltage antinode in the vicinity of the SQUID,
and having the higher frequency of the two modes.      

In the strong coupling regime (Fig. \ref{fig: density plot}c) the two lowest lines in the spectrum move to very low frequencies.
In this limit, the heat current modulation reappears again. 
In part, this effect is caused by the divergence of the Bose functions at low frequencies, 
which makes the relative contribution of these frequencies to the integral (\ref{eq: heat formula}) more significant.
In addition to that, the third hybrid mode with the frequency close to 
$\omega_{1,2}$ depends on the flux and also contributes to the modulation shown in Fig. \ref{fig: density plot}f.  

In the next three subsections we discuss each of the regimes introduced above in detail.

\subsection{Weak coupling regime}

In the weak coupling regime, the Hamiltonian of the combined system "resonators plus SQUID" 
can be approximately reduced to that of three coupled oscillators \cite{Koch2007,Photonic2019,Golubev2021},
\begin{eqnarray}
&& H =  \sum_{j=1,2}\hbar\Omega_r\left(a_j^\dagger a_j + \frac{1}{2}\right) 
+ \hbar\omega_J(\Phi) \left( b^\dagger b + \frac{1}{2} \right)
\nonumber\\ &&
-\, i\hbar g_1  (a_1^\dagger - a_1)(b^\dagger + b) - i\hbar g_2  (a_2^\dagger - a_2)(b^\dagger + b). 
\label{H}
\end{eqnarray}
Here $\Omega_r$ are the frequencies of the two lowest modes of the resonators shifted by presence of the capacitors $C_{c,j}$, 
\begin{eqnarray}
\Omega_r = \frac{\sqrt{\pi} \omega_j}{\sqrt{\pi + 4\omega_j Z_{0}C_{c,j}}},
\label{Omega_r}
\end{eqnarray}
$a_j$ are the ladder operators of the resonators, $b$ is the ladder operator of the SQUID, and
\begin{eqnarray}
\label{eq: g weak}
g_{j} = \sqrt{\frac{Z_{0}C_{c,j}^2\omega_j^3}{\pi(C_{c,1}+C_{c,2}+C)}}, 
\end{eqnarray}
are the coupling constants between the resonators and the SQUID. 
In the rest of the paper we consider the symmetric case and assume that the shifted frequencies (\ref{Omega_r}) are the same for both resonators.
This implies that the parameters $\omega_j$ and $C_{c,j}$ are not independent.
Eqs. (\ref{Omega_r}) and (\ref{eq: g weak}) have been derived by expanding 
the tangents in the resonator impedances (\ref{eq: bath impedance}) as
\begin{equation}
\tan x=\sum_{n=0}^{\infty} \frac{2 x}{\pi^{2}\left(n+\frac{1}{2}\right)^{2}-x^{2}},
\end{equation}
and keeping only the pole in the expansion with $n=0$.
Eqs. (\ref{H}-\ref{eq: g weak}) are valid at small coupling $g_j\ll\Omega_r$.
Below we provide more accurate condition for the weak coupling approximation, which also involves the damping rate of the resonators (\ref{condition}).

Determining the normal modes of the Hamiltonian (\ref{H}), we find that one of them
is independent of the magnetic flux because it is uncoupled from the SQUID. 
We call this mode "uncoupled", it has a voltage node in the vicinity of the SQUID
and in the weak coupling limit its frequency always equals to that of the shifted resonator mode, $\omega_{\rm unc}=\Omega_r$.
The frequencies of the two other modes are
\begin{eqnarray}
\omega_\pm = \sqrt{\frac{\Omega_r^2+\omega_J^2 \pm \sqrt{(\Omega_r^2-\omega_J^2)^2 + 16 (g_1^2+g_2^2) \Omega_r\omega_J}}{2}}.
\nonumber\\
\label{omega_pm}
\end{eqnarray} 
In Fig. \ref{fig: density plot}a these modes are clearly visible, while the uncoupled mode 
overlaps with $\omega_\pm$ due to the small value of the coupling constant $g$.

In the weak coupling limit the heat current (\ref{eq: heat formula}) strongly increases 
at values of the magnetic flux $\Phi_r$, which correspond to the resonance condition $\omega_J(\Phi_r)=\Omega_r$, 
and it almost vanishes away from this point (Fig. \ref{fig: density plot}d).
In Fig. \ref{fig: density plot}a  the SQUID frequency crosses the resonator frequencies at the flux value $\Phi_r\approx 0.41\Phi_0$.
In this case the heat modulation amplitude equals to the maximum value of the heat flux. 
To find the latter,
it is sufficient to consider the range of frequencies $\omega\sim\omega_J\sim\Omega_r$,
where one can accurately approximate  the transmission probability (\ref{eq: tau formula}) as follows 
\begin{equation}
\tau(\omega) =  
\frac{ \kappa^2 {g}_{1}^2 {g}_{2}^2  }
{\left| (\omega - \omega_J)( \omega - \nu_r )^2 - (g_1^2+g_2^2)(\omega-\nu_r)  \right|^2}.
\label{tau_app}
\end{equation}
Here we have introduced the complex frequency $\nu_r = \Omega_r - i\kappa/2$, where
\begin{eqnarray}
\kappa = \frac{4 R \Omega_r}{\pi Z_{0}}
\label{kappa_weak}
\end{eqnarray}
is the damping rate of the resonator modes.

The heat current (\ref{eq: heat formula}) with the approximate transmission probability (\ref{tau_app}) 
can be evaluated analytically. If the temperatures of the resistors
are sufficiently high, $k_BT_j \gtrsim |\omega_J-\Omega|$, we obtain
\begin{eqnarray}
J(\Phi) = \frac{2{g}_1^2{g}_2^2}{{g}_1^2+{g}_2^2}
\frac{\left(\frac{{g}_1^2+{g}_2^2}{\kappa}+\frac{\kappa}{2}\right)\hbar\Omega_r[N_2(\Omega_r)-N_1(\Omega_r)]}
{(\omega_J(\Phi)-\Omega_r)^2 + \left(\frac{{g}_1^2+{g}_2^2}{\kappa}+\frac{\kappa}{2}\right)^2}.
\nonumber\\
\label{eq: Dima heat}
\end{eqnarray}
The maximum of the heat flux is achieved at  the resonance condition $\omega_J(\Phi)=\Omega$,
\begin{eqnarray}
J_{\rm max} = \frac{2{g}_1^2{g}_2^2}{{g}_1^2+{g}_2^2}
\frac{\hbar\Omega_r[N_2(\Omega_r)-N_1(\Omega_r)]}{\frac{{g}_1^2+{g}_2^2}{\kappa}+\frac{\kappa}{2}},
\label{Jmax}
\end{eqnarray}
while the minimum occurs far away from the resonance, i.e. either at zero flux, $\Phi=0$
or at $\Phi=\Phi_0/2$. As we have discussed, at weak coupling one always has $J_{\min}\ll J_{\max}$.
Therefore, in this regime the modulation of the heat current is close to its maximum value,
\begin{eqnarray}
\Delta J = J_{\max} - J_{\min} \approx J_{\max}.
\label{dJ_weak}
\end{eqnarray}
In the symmetric case $g_1=g_2=g$ and at very weak coupling $g\ll \kappa/2$ one finds
\begin{eqnarray}
\Delta J = \frac{2g^2}{\kappa}\hbar\Omega_r[N_2(\Omega_r)-N_1(\Omega_r)].
\label{dJ1}
\end{eqnarray}
Thus, in this limit the heat modulation grows with the coupling strength as $g^2$.
In the limit $g\gg\kappa/2$ the modulation saturates at the value
\begin{eqnarray}
\Delta J = \frac{\kappa}{2} \hbar\Omega_r[N_2(\Omega_r)-N_1(\Omega_r)].
\label{dJ2}
\end{eqnarray}

In Eq. (\ref{eq: Dima heat}) we have ignored the contributions of the high frequency modes of the resonators
to the heat transport. Since in our model the SQUID angular frequency $\omega_J(\Phi)$ does not cross these modes,
in the weak coupling regime they give small contribution.

In Fig. \ref{fig: min_max J}  we plot the maximum and the minimum values of the heat current,
obtained numerically, as a function of the coupling constant $g$ and compare them with the approximate results.
We note that the expressions (\ref{Jmax}) and (\ref{dJ_weak}) well agree with the numerics in the weak coupling regime.

Finally, we provide more accurate condition under which the weak coupling expressions (\ref{eq: Dima heat}-\ref{dJ2}) are valid,
\begin{eqnarray}
\frac{2g^2}{\kappa}+\frac{\kappa}{2} \lesssim |\omega_J(0)-\Omega| \lesssim \frac{k_BT_j}{\hbar}.
\label{condition}
\end{eqnarray}

\subsection{Intermediate coupling regime}

In this section we consider the intermediate coupling regime $g_j\sim\Omega_r$. 
In this case, the expressions for the coupling constants $g_j$ (\ref{eq: g weak}) and for other parameters should be corrected.
To obtain the corrected expressions, we consider the classical Lagrangian of the system.
For simplicity, we consider fully symmetric setup and put $\omega_1=\omega_2=\omega_r$, $C_{c,1}=C_{c,2}=C_c$ and $g_1=g_2=g$.
We also define the effective capacitance of the $\lambda/4$-resonators, $C_r=\pi/4 Z_0\omega_r$,
and their effective inductances $L_r=4Z_0/\pi\omega_r$. Afterwards, the classical Lagrangian of the lowest modes of the two resonators
interacting with the SQUID is expressed as
\begin{eqnarray}
{\cal L} &=& \frac{\hbar^2}{8e^2}\sum_{j=1,2}\left(C_r\dot\varphi_j^2 - \frac{\varphi_j^2}{L_r}+C_c(\dot\varphi_j-\dot\varphi)^2\right)
\nonumber\\ &&
+\, \frac{\hbar^2}{8e^2}\left( C\dot\varphi - \frac{2eI_C(\Phi)}{\hbar}\varphi^2 \right).
\label{Lag}
\end{eqnarray}
Here $\varphi$ is the Josephson phase of the SQUID and
$\varphi_j$ are the phases describing the resonators. They are related to the electric potentials at the ends of the resonators,
adjacent to the coupling capacitors, $V_j$, as $\dot\varphi_j=2eV_j/\hbar$. Diagonalizing the Lagrangian (\ref{Lag}),
we obtain the corrected expressions for the angular Josephson frequency $\omega_J$, for the angular frequencies of the resonator modes $\Omega_r$ 
and for the coupling constant $g$:
\begin{eqnarray}
\omega_J(\Phi)=\sqrt{\frac{2e I_C(\Phi) (C_r+C_c)}{\hbar( CC_c + (C+2C_c)C_r)}},
\label{omega_J1}
\end{eqnarray}
\begin{eqnarray}
\Omega_r  = \omega_r\sqrt{\frac{(C+2C_c)C_r}{CC_c + (C+2C_c)C_r}},
\label{Omega_r1}
\end{eqnarray}
\begin{eqnarray}
g =\omega_r \sqrt{\frac{C_c^2C_r}{4(C_r+C_c)(CC_c + (C+2C_c)C_r)} }.
\label{g1}
\end{eqnarray}
In the limit $C_c\ll C_r$ these expressions match the weak coupling formulas given in the previous section.
In addition, if the resistance $R$ approaches $Z_0$ one should use more accurate expression for the damping rate,
\begin{eqnarray}
\kappa = \frac{4RZ_0\omega_r}{\pi|Z_0^2-R^2|}.
\label{kappa_1}
\end{eqnarray}
With these corrections, Eq. (\ref{eq: Dima heat}) approximately describes the heat current in the intermediate coupling regime.
The frequencies of the eigenmodes of the coupled system in the limit $R\to 0$ are 
\begin{eqnarray}
\omega_{\rm unc} = \omega_r\sqrt{\frac{C_r}{C_r+C_c}}
\label{omega_unc}
\end{eqnarray}
for the mode uncoupled from the SQUID, and
\begin{eqnarray}
\omega_\pm = \sqrt{\frac{\Omega_r^2+\omega_J^2 \pm \sqrt{(\Omega_r^2-\omega_J^2)^2 + 32 g^2 \omega_J^2}}{2}}
\label{omega_pm_2}
\end{eqnarray}  
for the two hybrid modes. Note that the interaction term in this equation slightly differs from that in Eq. (\ref{omega_pm}).
As expected, in the limit $C_c\ll C_r$ these expressions meet those of the previous section.

The main difference between the weak and the intermediate
coupling regimes is in the growing value of the minimum heat current. Assuming that $J_{\min}=J(0)$, 
from Eq. (\ref{eq: Dima heat}) one finds the modulation in the form
\begin{eqnarray}
\Delta J =
\frac{g^2(\omega_J(0)-\Omega_r)^2\hbar\Omega_r[N_2(\Omega_r)-N_1(\Omega_r)]}
{\left((\omega_J(0)-\Omega_r)^2 + \left(\frac{2g^2}{\kappa}+\frac{\kappa}{2}\right)^2\right)\left(\frac{2g^2}{\kappa}+\frac{\kappa}{2}\right)}.
\label{dJ_int}
\end{eqnarray} 
This modulation amplitude significantly drops if
\begin{eqnarray}
\frac{2g^2}{\kappa}+\frac{\kappa}{2} \gtrsim |\omega_J(0)-\Omega_r|,
\label{condition3}
\end{eqnarray}
i.e. as soon as the coupling can no longer be considered weak. 

To illustrate these points, 
in Fig. \ref{fig: density plot}b we show the transmission probability $\tau(\omega,\Phi)$ in the intermediate coupling regime $g\sim\Omega_r$.
The flux-independent line at $f\approx 4.4$ GHz corresponds to the uncoupled mode (\ref{omega_unc}).
The lines corresponding to hybrid modes (\ref{omega_pm_2}) are well separated at all values of magnetic flux.
This is why the dependence of the heat current on $\Phi$ becomes weak and does not exhibit a resonance peak (Fig. \ref{fig: density plot}e). This, in turn, suppresses
the heat current modulation, as is evident from Fig. \ref{fig: min_max J}.

\subsection{Strong coupling regime}

In the strong coupling limit the hybrid mode $\omega_-(\Phi)$ (\ref{omega_pm_2}) and the uncoupled mode $\omega_{\rm unc}$ (\ref{omega_unc}) move to 
low frequencies, where they merge and form a broad peak in $\tau(\omega,\Phi)$, which depends on $\Phi$.
The mode $\omega_+(\Phi)$ becomes isolated, with pronounced dependence on the magnetic flux due the strong coupling to the SQUID.
This behavior of the transmission peaks is visible in Fig. \ref{fig: density plot}c.
These effects lead to the reappearing of the heat current modulation in the strong coupling limit.

Formal boundaries of the strong coupling limit, in which one can derive approximate analytic expressions, are 
\begin{eqnarray}
C_c\gg \max\{C_r,C,L_r/R^2\}.
\label{condition2}
\end{eqnarray} 
In the limit $C_c\gg \max\{C_r,C\}$ Eqs. (\ref{omega_J1}-\ref{g1}) become
\begin{eqnarray}
\omega_J(\Phi)=\sqrt{\frac{2e\hbar I_C(\Phi)}{C+2C_r}},
\end{eqnarray}
\begin{eqnarray}
\Omega_r  = \omega_r\sqrt{\frac{2C_r}{C+2C_r}},\;\;\;
g =\omega_r \sqrt{\frac{C_r}{4(C+2C_r)} }.
\end{eqnarray}

\begin{figure}
\includegraphics[width=\columnwidth]{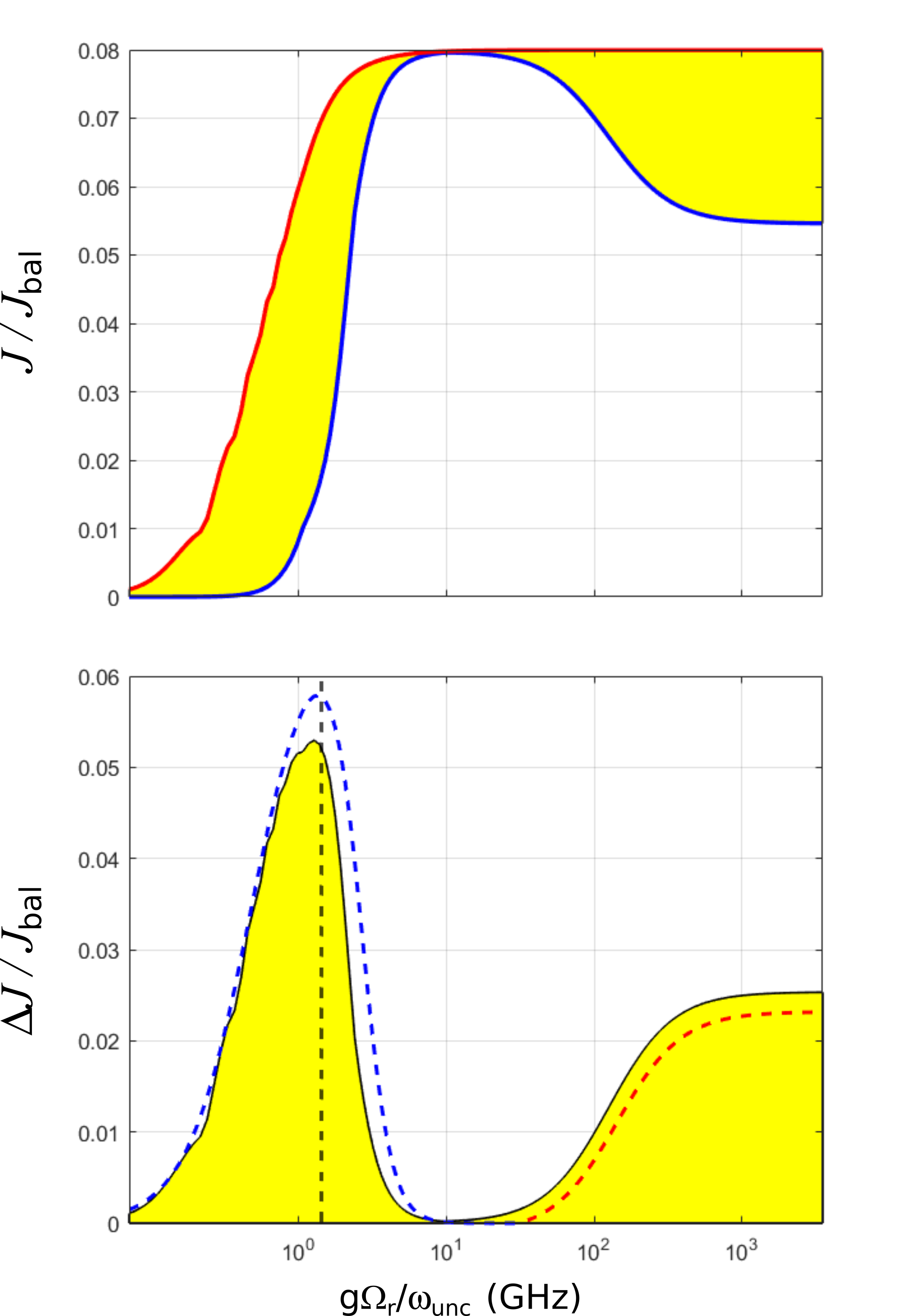}
\caption{Top panel: the maximum (red line) and the minimum (blue line) photonic heat power (\ref{eq: heat formula}) 
as a function of the coupling strength $g$ (\ref{g1}). 
Bottom panel: the heat power modulation $\Delta J = J_{\max} - J_{\min}$ as a function of $g$. 
In both plots we have normalized the heat power by ballistic heat conduction, $J_{\rm bal}=(\pi k_B^2/12\hbar)(T_2^2-T_1^2)$,
which follows from Eq. (\ref{eq: heat formula}) if $\tau(\omega)\equiv 1$.
For visual clarity, the horizontal axis shows the re-scaled coupling strength $g\Omega_r/\omega_{\rm unc}$.
In the bottom panel we also show the analytical approximations based on Eq. (\ref{dJ_int}) (dashed blue line) and 
Eqs. (\ref{Jl},\ref{eq: heat high 2}) (dashed red line). 
Vertical black dashed line shows the predicted optimal coupling strength $g\approx\sqrt{\kappa\Omega_r/2}$, which follows
from the condition (\ref{condition3}) assuming $g\gtrsim\kappa/2$.}
        \label{fig: min_max J}
\end{figure}

Furthermore, at low frequencies $\omega\ll \omega_r$ one can
approximate the impedances of the resonators (\ref{eq: bath impedance}) as 
$Z_1(\omega)=Z_2(\omega)=-i\omega L_r+R$.
In this limit, at small capacitance of the SQUID, $C\ll L_r/R^2$ and for $C_c\gg L_r/R^2$  
the transmission probability at low frequencies acquires the form of a non-Lorentzian peak,
\begin{eqnarray}
\tau(\omega)=
\frac{4R^2\omega^2}{\left(  \frac{2eI_C(\Phi)}{\hbar} (\omega^2 L_r^2 + R^2) + 2\omega^2 L_r  \right)^2 + 4R^2\omega^2 }.
\label{tau_strong}
\end{eqnarray}
This peak has a maximum at frequency $\omega_{\max}\sim R/L_r$.
The contribution of the low frequency peak to the heat flux can be evaluated analytically for temperatures 
$T_1,T_2\gtrsim {\hbar R}/{k_BL_r}$. 
In this case, in Eq. (\ref{eq: heat formula}) one can make the low frequency approximation for the Bose functions
\begin{eqnarray}
\hbar\omega[N_2(\omega)-N_1(\omega)] \to k_B(T_2-T_1).
\end{eqnarray}
After that, Eq. (\ref{eq: heat formula}) with the transmission probability (\ref{tau_strong}) leads to the
contribution to the heat flux coming from the low frequency peak in the form:
\begin{eqnarray}
J_l(\Phi) = \frac{\kappa k_B(T_2-T_1)}{4(1+A(\Phi))(2+A(\Phi))}.
\label{Jl1}
\end{eqnarray}
Here we have defined the flux dependent dimensionless parameter 
\begin{eqnarray}
A(\Phi)=\frac{\pi e Z_0I_C(\Phi)}{\hbar\omega_r} = \frac{\pi^2}{8} \frac{L_r}{L_J(\Phi)}.
\end{eqnarray}
One can work out even more accurate approximation for the low frequency contribution,
\begin{eqnarray}
J_l(\Phi) = \frac{\kappa k_B(T_2-T_1) }{4(2+A(\Phi))}\frac{A(\Phi)\kappa^2+4(1+A(\Phi))\omega_{\rm unc}^2}{A(\Phi)(1+A(\Phi))\kappa^2+4\omega_{\rm unc}^2}.
\nonumber\\
\label{Jl}
\end{eqnarray}
This result can be extended to the intermediate coupling regime, i.e. to the values of $C_c$ smaller than the condition (\ref{condition2}) requires.

The contribution of the mode $\omega_+$ (\ref{omega_pm_2}) to the heat current can be estimated as 
\begin{equation}
\label{eq: heat high 2}
J_+(\Phi) = \frac{\kappa}{4}\hbar \omega_+(\Phi) [ N_2(\omega_+(\Phi)) - N_1(\omega_+(\Phi))], 
\end{equation}
where the frequency $\omega_+(\Phi)$ is given by Eq. (\ref{omega_pm_2}). In the limit $C_c\gg \max\{C_r,C\}$, where $g^2=\Omega_r^2/8$, 
this frequency acquires a simple form
\begin{eqnarray}
\omega_+(\Phi)&=&\sqrt{\Omega_r^2+\omega_J^2(\Phi)}
\nonumber\\
&=&\sqrt{ \omega_r^2 \frac{2 C_r}{C + 2 C_r} + \frac{2eI_C(\Phi)}{\hbar (C + 2 C_r)}}.
\end{eqnarray}
The total heat current takes the form
\begin{eqnarray}
J(\Phi) = J_l(\Phi) + J_+(\Phi) + J_{\rm bg}(\Phi),
\end{eqnarray}
where $J_{\rm bg}(\Phi)$ is the background contribution coming from the modes with frequencies higher than $\omega_+$.
Interestingly, in the strong coupling regime the heat current has a maximum at $\Phi=0.5\Phi_0$ and minimum --- at $\Phi=0$,
see Fig. \ref{fig: density plot}f.

In Fig. \ref{fig: min_max J} we observe the reappearance of the heat current modulation at strong coupling.
For the chosen parameters the modulation predominantly comes from the term $J_+(\Phi)$ (\ref{eq: heat high 2}), 
although the low frequency part $J_l(\Phi)$ also gives significant contribution.  In the limit $C_c\to \infty$
and for $k_BT_{1,2}\gtrsim \omega_+(\Phi)$ the modulation approaches the limiting value
\begin{eqnarray}
\Delta J = \left[\frac{A(1+A)}{(1+A)(2+A)}
+ \frac{\hbar^2\omega_+^2}{6k_B^2 T_1T_2}\right]
\frac{\kappa k_B(T_2-T_1)}{8},
\end{eqnarray}
where both $A$ and $\omega_+$ are taken at $\Phi=0$.

\section{Conclusion}

In conclusion, we have studied photonic heat transport through a SQUID coupled to the two resonators and two resistors.
By tuning the critical current of the SQUID with magnetic flux, one can control the heat power transmitted from the hot
resistor to the cold one. This device can be used as a heat valve provided its' parameters are chosen properly. 
We study the performance of the heat valve depending on the coupling strength between the resonators and the SQUID.
We find that the main parameter characterizing the performance of such device, namely, the amplitude of modulation of
the photonic heat power, non-monotonously varies with the coupling strength. The modulation grows with the coupling
strength in the weak coupling regime, then significantly drops at the intermediate coupling and, finally, it reappears
again in the strong coupling limit. 
This unusual behavior is explained by the resonant nature of the heat transport in the system. Indeed, at weak coupling the heat
flows through the device only at magnetic flux values corresponding to the resonance condition $\omega_J(\Phi)=\Omega_r$
and drops to zero away from these values. As a result, the dependence of the heat power on the magnetic flux, $J(\Phi)$, is
given by a periodic set of narrow peaks. At the intermediate coupling these peaks become broader and eventually overlap, thus reducing the
heat modulation. 
The optimal performance of the heat valve is achieved at the boundary 
between the weak and the intermediate coupling regimes. Our results can help to optimize the design of the low temperature heat
valves based on superconducting circuit components.

\acknowledgements
We acknowledge the support by the Academy of Finland Centre of Excellence program (project 312057).
We would also like to acknowledge very helpful discussions with Jukka P. Pekola.

\twocolumngrid
\bibliography{references}
\onecolumngrid

\end{document}